\documentclass{elsart}
\usepackage{natbib}
\usepackage{graphicx}
\usepackage{amssymb}

\begin{document}

\begin{frontmatter}


\title{Power-law versus exponential distributions of animal group sizes}

\author{Hiro-Sato Niwa}
\ead{Hiro.S.Niwa@fra.affrc.go.jp}

\address{Behavioral Ecology Section,
National Research Institute of Fisheries Engineering,
Hasaki, Ibaraki 314-0421, Japan}

\begin{abstract}
There has been some confusion concerning the animal group-size:
an exponential distribution was deduced by maximizing the entropy;
lognormal distributions were practically used;
a power-law decay with exponent~{3/2} was proposed in physical analogy to aerosol condensation.
Here I show that the animal group-size distribution follows a power-law decay with exponent 1, and is truncated at a cut-off size
which is the expected size of the groups an arbitrary individual engages in.
An elementary model of animal aggregation based on binary splitting and coalescing on contingent encounter is presented.
The model predicted size distribution holds for various data from pelagic fishes and mammalian herbivores in the wild.
\end{abstract}


\end{frontmatter}

\section{Introduction}
It is hard to find animals in nature that do not aggregate
(school, herd, swarm, or flock).
Despite the universality of aggregation and its ecological and economic importance for the estimation of wildlife abundance,
the statistical question about animal group-size has been involved in much confusion.
Theoretically the Gibbs-Boltzmann (exponential) distribution has been proposed for the animal group-size distribution  by applying a maximum entropy principle \citep{Okubo86},
whereas practically the lognormal distribution has been used in fisheries \citep{Matsuishi-etal93}.
Only lately, it has been suggested that power-law distributions may be
quite generic \citep{Bonabeau-Dagorn95}.
\citet{Anderson81}, a pioneer in statistical research into group formation, proposed a stochastic dynamic equation for the size of fish school, in which the possibility of power laws was already presented but not exploited.
\citet{Bonabeau-Dagorn95} presented a model of animal aggregation inspired by a physical model of particle aggregation \citep{Takayasu89},
and predicted that the group-size distributions follow a power-law decay with exponent~{3/2}.
The empirically determined power-law size distributions of fish schools have exponents in the range from 0.7 to 1.8 (Niwa, 1998; Bonabeau {\it et al.}, 1999).

$N$ animals moving together form an $N$-sized group.
The group-size distribution $W(N)$ is proportional to the observed number of $N$-sized groups.
Power-law distributions, $W(N) \propto N^{-\beta}$, have somewhat unusual properties.
They do not have a well-defined mean when $\beta\leq 2$.
Fat-tailed group-size distributions are necessarily truncated at a cut-off size because the population is finite,
but truncated power law must be distinguished from purely rapidly decreasing ones, as they exhibit specific properties,
e.g., a violation of the central-limit theorem (Mantegna \& Stanley, 1995; Axtell, 2001).
In order to investigate the possibility of power-law regimen in the group-size distribution,
let us reconsider some existing data (Table~\ref{tab1}) in terms of the population distribution
$P(N)=NW(N)$,
which is proportional to the ratio of animal population in $N$-sized groups to total population.

Suppose that we have a data set arranged in a frequency distribution with $n$ classes.
Let $W_i \Delta N$ be the frequency of animal groups observed in the
$i$-th class (where $\Delta N$ is the class width);
the class mark of group size, $N_i$, is the midpoint of the class.
$W_i$ then represents the distribution density of group sizes.
The number of individuals associated with groups of the $i$-th class is
given by $N_i W_i \Delta N$
 (denoted by $P_i \Delta N$).
$P_i$ then represents the distribution density of population.
Shown in Fig.\ref{fig:1} is a semilogarithmic plot of the population distribution of pelagic fishes.
Surprisingly, I find that the scaled population distributions appear to collapse onto a single curve, an exponentially decreasing function.
It implies that
(i)~the population distribution has a well-defined mean
\begin{equation} \label{eqn:pmean}
 \langle N \rangle_P =
 \left.
  \sum_{i=1}^n N_i P_i
 \right/
  \sum_{i=1}^n P_i
\end{equation}
which is hereafter referred to as $P$-mean;
(ii)~the school-size distribution displays robust scale-invariant behavior with the power-law index $\beta =1$;
and
(iii)~the power-law distribution of school sizes is truncated at the cut-off size equal to $P$-mean.
$P$-mean is the average of the population distribution among group sizes, i.e., the expected size of the groups in which an arbitrary individual engages.
The population distribution is the crossover function from power-law to exponential decay.

\section{The Model}
The empirical finding for fish school size suggests that the population distributions among school sizes are identical for fishes regardless of large diversities in behavioral and ecological conditions.
I propose a simple stochastic-differential-equation model accounting for group-size statistics.
Assume binary splitting independent of group size (breakup rate $p$) and coalescing on contingent encounter.
Trace the size change of the group that a certain individual (named ``A'') rides.
The group size at time $t$ is denoted by $N_{\mbox{\scriptsize A}}(t)$.
A discrete countable number of group size is replaced by a continuous group size, providing that there is a sufficiently large reservoir of individuals.
Rewriting the difference (the size change 
$\Delta N_{\mbox{\scriptsize A}}$
for a finite time interval
$\Delta t$)
to the differential of the group size for an infinitesimal time interval
$\mbox{d}t$,
\begin{equation}
 \Delta N_{\mbox{\scriptsize A}}(t)
  =
  N_{\mbox{\scriptsize A}}(t+\Delta t)-N_{\mbox{\scriptsize A}}(t)\;
  \to\;
  \mbox{d}N_{\mbox{\scriptsize A}}(t)
\end{equation}
gives the Ito stochastic differential equation for group size:
\begin{equation} \label{eqn:SDE}
 \mbox{d}N_{\mbox{\scriptsize A}}(t)
 =
 -\frac p2 \left(N_{\mbox{\scriptsize A}}(t) - \langle N \rangle_P \right) \mbox{d}t
 +
 \sigma \left(N_{\mbox{\scriptsize A}}(t)\right) \mbox{d}B(t)
\end{equation}
where
$\mbox{d}B$ is Wiener noise term of normal distribution with zero mean and variance $\mbox{d}t$, and $\sigma \left(N_{\mbox{\scriptsize A}}\right)$
is the standard deviation of group-size changes.
Fluctuations of the group size in aggregation-breakup processes,
$\sigma \left(N_{\mbox{\scriptsize A}}\right)$,
are numerically determined as follows:

Assume for simulations that there are $s$ sites, coarse-grained zones of space, on which $\Phi$ individuals move.
The number of individuals, $\Phi$, is conserved.
Each site is either empty or occupied by one group.
At each discrete time step, all groups move towards a randomly selected site.
They may move to any site with equal probability.
This corresponds to the mean-field theory.
When $M$- and $N$-sized groups happen to move to the same site, they aggregate to form an $(M+N)$-sized group.
Each group with a size greater than or equal to 2 splits into a pair of groups with the probability $\hat{p}$ at each time step.
It is assumed that the probability $\hat{p}$ for a group to split is independent of its size and that the sizes of splitting groups are uniformly distributed:
a probability for an $N$-sized group to split into $M$- and $(N-M)$-sized groups is represented by
\begin{equation} \label{eqn:split-prob}
 K_{\mbox{\scriptsize b}}(N|M,N-M)
 =
 K_{\mbox{\scriptsize b}}(N)
 =
  \frac{\hat p}{N-1}
\end{equation}
for $N \geq 2$.
Then, the expected decrement and increment of size of the group that a certain individual ``A'' rides are
$\hat{p} N_{\mbox{\scriptsize A}}/2$
and
$\Phi/s$
(i.e., coarse-grained spatial population density)
at each time step, respectively.
Monte-Carlo simulations of animal-group aggregation show that fluctuations of the group size in aggregation-breakup processes exponentially depend on the size (Fig.\ref{fig:2}), and give
\begin{equation} \label{eqn:variance}
 \sigma^2 (N)
 =
 2D \exp \left(\frac N{\langle N\rangle_P} \right)
\end{equation}
where $D$ denotes the size of fluctuations.
The same results can be obtained with another model for a group to split: the size distribution of splitting groups is binomial.
\begin{equation}
 K_{\mbox{\scriptsize b}}(N|M,N-M)
 =
 \frac{\hat p}{2^N -2}
 {N \choose M}
\end{equation}

Let the domain of the size $N$ be real numbers.
The aggregation-breakup dynamics [eqn (\ref{eqn:SDE})] is supposed to be symmetric about $N=0$.
By performing the change of variable \citep{Richmond01}
\begin{equation} \label{eqn:vchange}
 \frac{\mbox{d}x}{\mbox{d}N}
  =
  \frac{\sqrt{p/2}}{\sigma (N)}
\end{equation}
eqn (\ref{eqn:SDE}) becomes the Langevin equation for a Brownian particle moving in a potential $U(x)$:
\begin{equation} \label{eqn:langevin}
 \mbox{d}x(t)
  =
  -\frac{\partial U}{\partial x}\mbox{d}t+ \sqrt{\frac p2} \mbox{d}B(t)
\end{equation}
where
\begin{equation} \label{eqn:potential}
 U(x)
  =
  -\frac p4
  \ln \left(1-\frac{x}{\langle N\rangle_P \sqrt{p/D}}\right)
  \left[
   1-
   \frac{p\langle N\rangle_P^2}D
   \left(1-\frac{x}{\langle N\rangle_P \sqrt{p/D}}\right)^2
 \right]
\end{equation}
The induced potential is schematically depicted in Fig.\ref{fig:3}.
If the potential $U(x)$ is continuously differentiable, the following relation is obtained:
\begin{equation} \label{eqn:fluct-dissip}
p \langle N\rangle_P^2 = D
\end{equation}
Therefore the breakup rate $p$, which corresponds to the dissipation in the stochastic differential equation, and the noise $D$ cannot be independent.
The numerical results indicate that this fluctuation-dissipation relation is already satisfied (Fig.\ref{fig:2}).

\section{Results}
The stationary solution of eqn (\ref{eqn:SDE}) reads the probability for individual ``A'' to be found in $N$-sized groups, which is proportional to the stationary population distribution among group sizes, $P(N)$.
The distribution $P(N)$ has the desired form and the stationary group-size distribution $W(N)\; [=N^{-1}P(N)]$ is rigorously written as
\begin{equation} \label{w-distr}
 W(N)
 \propto
 N^{-1}
  \exp \left[
  -\frac N{\langle N \rangle_P}
  \left(
  1-\frac{e^{-N/\langle N \rangle_P}}2
   \right)
 \right]
\end{equation}
where eqn (\ref{eqn:variance}) is adopted for the variance with eqn (\ref{eqn:fluct-dissip}).
There is no parameter to be adjusted, since $P$-mean is obtained directly from the data
[because of the fluctuation-dissipation relation, eqn (\ref{eqn:fluct-dissip}), parameters $p$ and $D$ do not explicitly appear in the group-size distribution].
The self-consistency in the average of population distribution
\begin{equation} \label{eqn:s-consist}
 \frac{
  \displaystyle{
  \int_0^{\infty} q \exp
  \left[-q \left( 1-e^{-q}/2 \right) \right]\mbox{d}q
  }
  }{
  \displaystyle{
  \int_0^{\infty} \exp
  \left[-q \left( 1-e^{-q}/2 \right) \right]\mbox{d}q
  }
  }
  = 1
\end{equation}
is numerically certified by using $2^8$-digit precision for internal computations,
where $q=N/\langle N \rangle_P$.

The model accounting for group-size statistics presented here is consistent with a number of numerical experiments.
Fig.\ref{fig:4} represents the group-size distribution obtained from simulations.
The model is tested with empirical data for the fish school-size distributions and a close agreement is found (Fig.\ref{fig:5}).
The normalization factor is given by
\begin{equation} \label{eqn:normal-factor}
 \left[
  \int_0^{\infty} \exp
  \left[-q \left( 1-e^{-q}/2 \right) \right]\mbox{d}q
\right]^{-1}
 = 0.881237
\end{equation}

The size distribution $W(N)$ follows a power law with exponent $\beta =1$ to a cut-off size $\langle N \rangle_P$, i.e., a well-defined mean of the associated distribution $P(N)$ which is exponentially decaying function.
Naturally, a well-defined mean does not exist for the group-size distribution.
Moreover, a coarse-grained description of the group aggregation suggests that $P$-mean varies in proportion as the number of individuals or the spatial population density $\rho$, and inversely as the breakup rate:
\begin{equation} \label{eqn:popsize-Pmean}
\langle N \rangle_P \propto \frac{\rho}p
\end{equation}
which is numerically approved (Fig.\ref{fig:6}).
The group-size distribution, therefore, follows a universal scaling law, i.e., it decays as a power law with an exponential truncation controlled not only by the system size (number of individuals) but also the breakup rate.
I indeed find a collapse of the empirical data onto a single curve as well as the collapse of the numerical results.
Eqn (\ref{eqn:popsize-Pmean}) is utilized for data normalization (Table~\ref{tab1}).

\section{Discussion}
I have reconsidered some existing data on pelagic fishes
and found identical statistical signatures:
the school-size distribution follows a power-law decay with the exponent 1 up to a cut-off size which is given by $P$-mean.
The rationale behind this robust scaling invariance is the exponential fluctuation of school-size change.
I have proposed an Ito stochastic differential equation governing the evolution of the group size that a certain individual rides, which is quite different approach from Anderson's (1981) model based on a stochastic dynamics for the change of the number of individuals in a certain group.
The model predicts not only the power-law behavior observed in nature, but also the deviation from pure power-law towards exponential decay.
The similarity between the empirical distribution and the distribution obtained from the model is striking, though there is no fitting parameter.
The exponential fluctuation of group size is the essential ingredient of underlying aggregate-breakup dynamics that influence animal group size.
If the size fluctuation $\sigma^2 (N)$ were proportional to the group size $N$, the group-size distribution would follow a power-law decay with exponent 3/2, as was suggested by \citet{Bonabeau-Dagorn95}.

The interacting group systems introduced here have a marked feature in contrast to physical systems in thermal equilibrium:
the aggregate-breakup process in animal-group systems does not allow for detailed balance as below;
while \citet{Okubo86} made the detailed balance assumptions in order to obtain the group-size distribution.

The Smoluchowski rate equation is the alternative equation governing the
group-size distribution
(Gueron \& Levin, 1995; Gueron, 1998; Durrett {\it et al.}, 1999),
and is equivalent to the following master equation for associated distribution of population \citep{Niwa98}:
\begin{eqnarray} \label{eqn:pauli}
\lefteqn{
 P(N,t+\Delta t)-P(N,t)}\nonumber\\
 && \mbox{ } =
  -P(N,t) \sum_{M=1}^\Phi m(N,M;t)
  +\sum_{M=1}^\Phi P(M,t) m(M,N;t)
\end{eqnarray}
with the transition probability for a specific individual which engages in a group of size $N$ to pass into a group of size $M$ in a finite time interval $\Delta t$:
\begin{eqnarray} \label{eqn:transition-rate}
\lefteqn{
 m(N,M;t)}\nonumber\\
 && \mbox{ } =
  K_{\mbox{\scriptsize b}}(N|M,N-M) \frac MN
  + P(M-N,t)
  \frac{K_{\mbox{\scriptsize a}}(N,M-N|M)}{M-N}
\end{eqnarray}
where $K_{\mbox{\scriptsize a}}$ denotes the probability for an $N$-sized group to join with an $M$-sized group in a finite time interval $\Delta t$.
Whenever two groups meet they are supposed to join,
which implies
$K_{\mbox{\scriptsize a}} = s^{-1}$
for a coarse-grained model of animal-group system.
Assume the detailed balance condition:
there are as many transitions per $\Delta t$ from $M$-sized groups to $N$-sized groups as from $N$ to $M$ by the inverse process.
The following recurrence formula for stationary solution is then obtained
[$K_{\mbox{\scriptsize b}}$ is given by eqn (\ref{eqn:split-prob})]
\begin{equation}
 P(N) = P(N-1) N \frac{P(1)}{s{\hat p}}
\end{equation}
which leads to an explicit solution for group-size distribution
\begin{equation} \label{eqn:d-b-W}
 W(N) = \frac{W(1)}{e^{\mu} \pi_{N-1,\mu}}
\end{equation}
where $\pi_{N,\mu}$ is the Poisson distribution with mean
$\mu = s{\hat p}/W(1)$.
The result contradicts the simulated group-size distribution;
besides, eqn (\ref{eqn:d-b-W}) violates the conservation law of population.
The detailed balance conditions therefore do not hold in the
aggregate-breakup systems,
or else we may have another deterministic equation which is consistent
with micro-reversibility.

The model developed here can apply to a wide spectrum of cases on animal species.
Fig.\ref{fig:7} shows the herd-size distributions for mammalian herbivores;
these distributions are much better fitted by a power law than by an exponential decay, as was suggested by \citet{Okubo86}.
We may however notice that the empirical data are not in perfect agreement with eqn (\ref{w-distr}).
It may be explained by biases in the data.
Another possibility bringing a subtle difference between data and theory is that
the uniform-breakup assumption (splitting probability independent of group size) is no longer valid.

There is yet another possibility.
This subtle difference may be caused by the fluctuation-dissipation relation reduced above.
It can be generalized as follows:
\begin{equation} \label{eqn:g-fluct-dissip}
p \langle N\rangle_P^2 = x_0^2 D
\end{equation}
where $x_0^2 \geq 1$, and potential $U(x)$ is no longer differentiable at $x=0$ but continuous (Fig.\ref{fig:3}).
The population distribution among group sizes is then modified into
\begin{equation} \label{eqn:g-p-distr}
 P(N)
  \propto
  \exp \left[
	-\frac N{\langle N \rangle_P}
	\left(
	 1-\frac{p \langle N \rangle_P^2}{2D} e^{-N/\langle N \rangle_P}
       \right)
      \right]
\end{equation}
which guarantees the self-consistency in the average of distribution as well as eqn (\ref{eqn:s-consist}).
The modified size distribution by adopting eqn (\ref{eqn:g-fluct-dissip}) matches better the data.
Yet, the modified distribution [eqn (\ref{eqn:g-p-distr})] can apply to schools of sardinellas caught in the up-welling areas \citep{Bonabeau-etal99}, and the agreement between the empirical distribution and the model's prediction is remarkable (Fig.\ref{fig:8}).

\newpage
\begin{flushleft}
{\bf Figure legends}
\end{flushleft}

\begin{figure}[p]
 \centering
 \includegraphics[width=10cm]{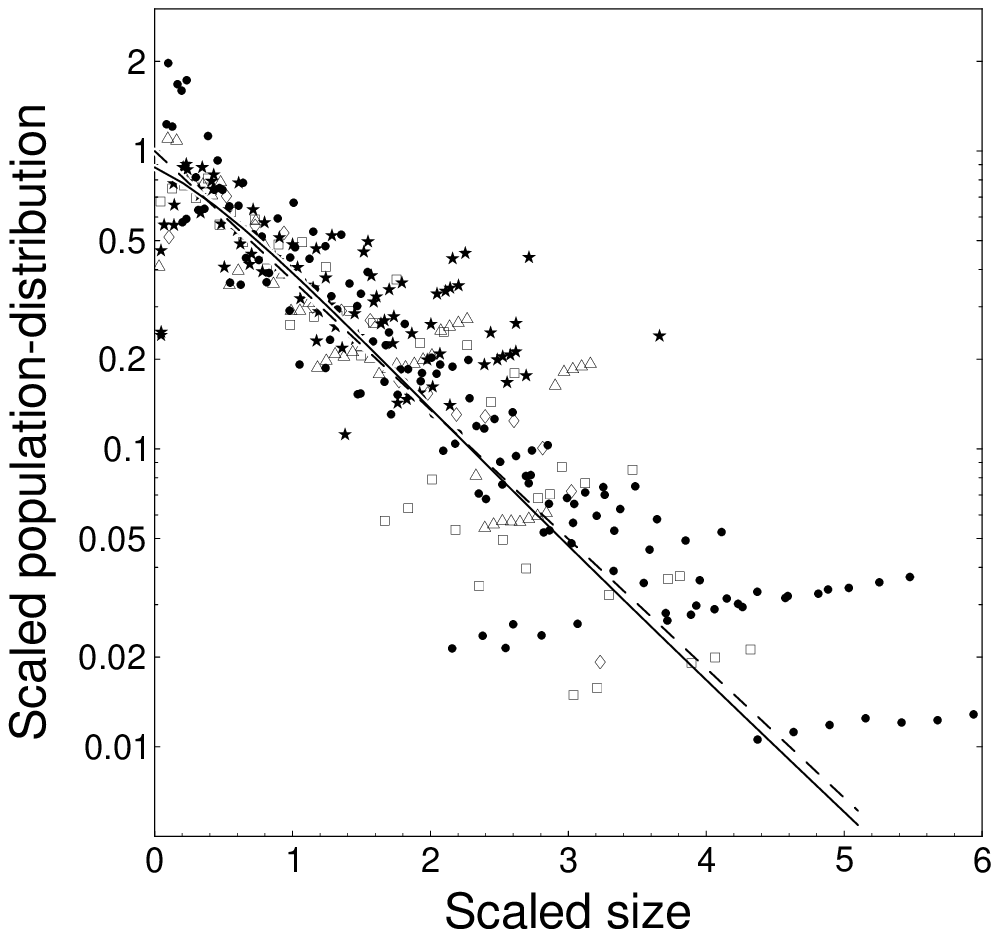}
 \caption{}
 \label{fig:1}
\end{figure}

\begin{figure}[p]
 \centering
 \includegraphics[width=10cm]{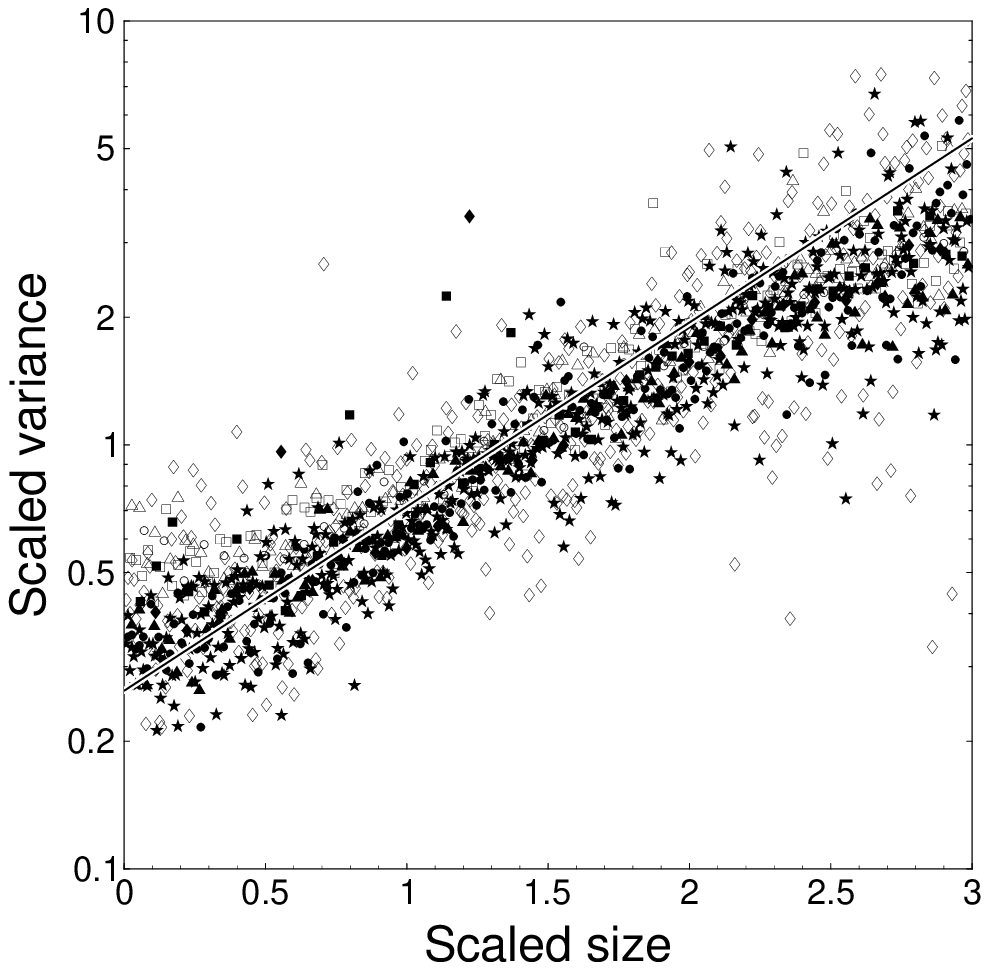}
 \caption{}
 \label{fig:2}
\end{figure}

\begin{figure}[p]
 \centering
 \includegraphics[width=10cm]{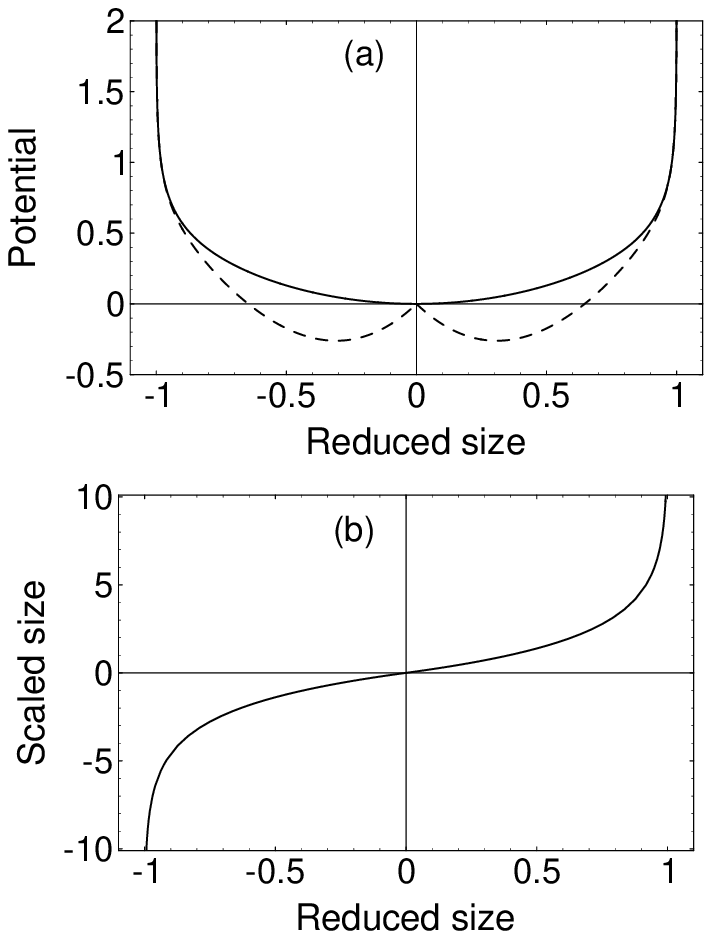}
 \caption{}
 \label{fig:3}
\end{figure}

\begin{figure}[p]
 \centering
 \includegraphics[width=10cm]{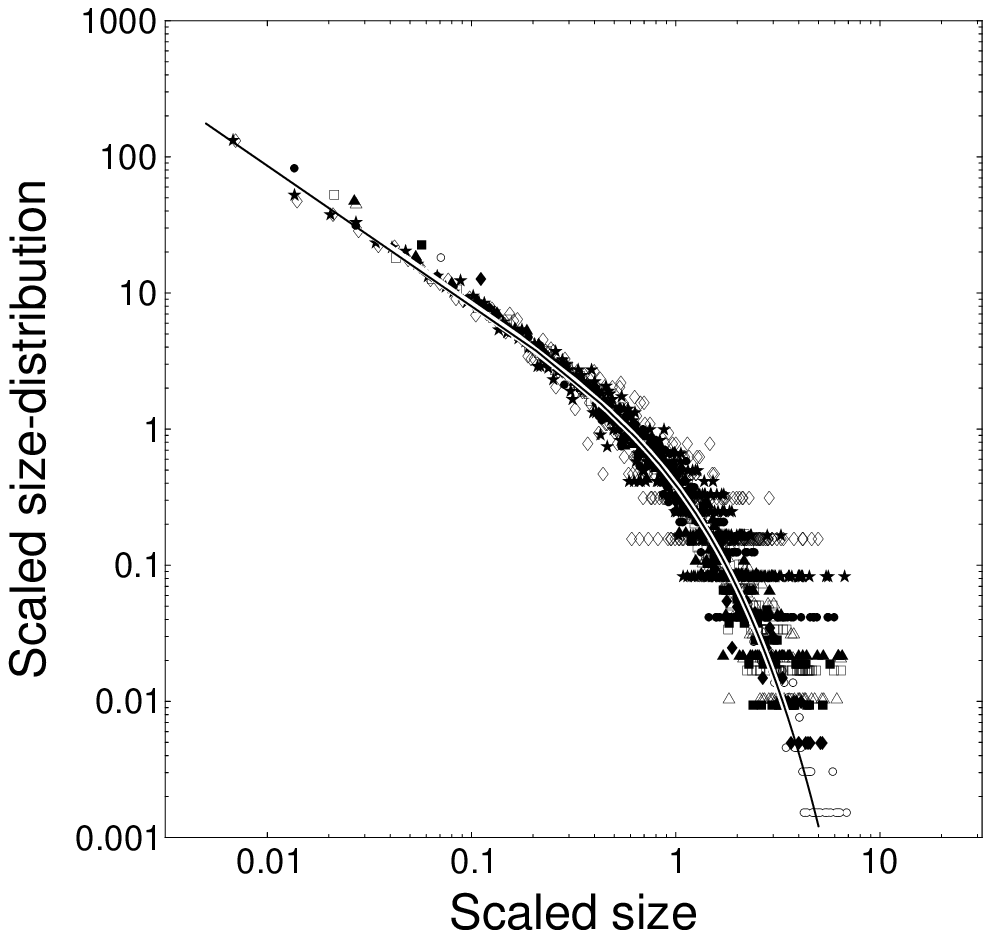}
 \caption{}
 \label{fig:4}
\end{figure}

\begin{figure}[p]
 \centering
 \includegraphics[width=10cm]{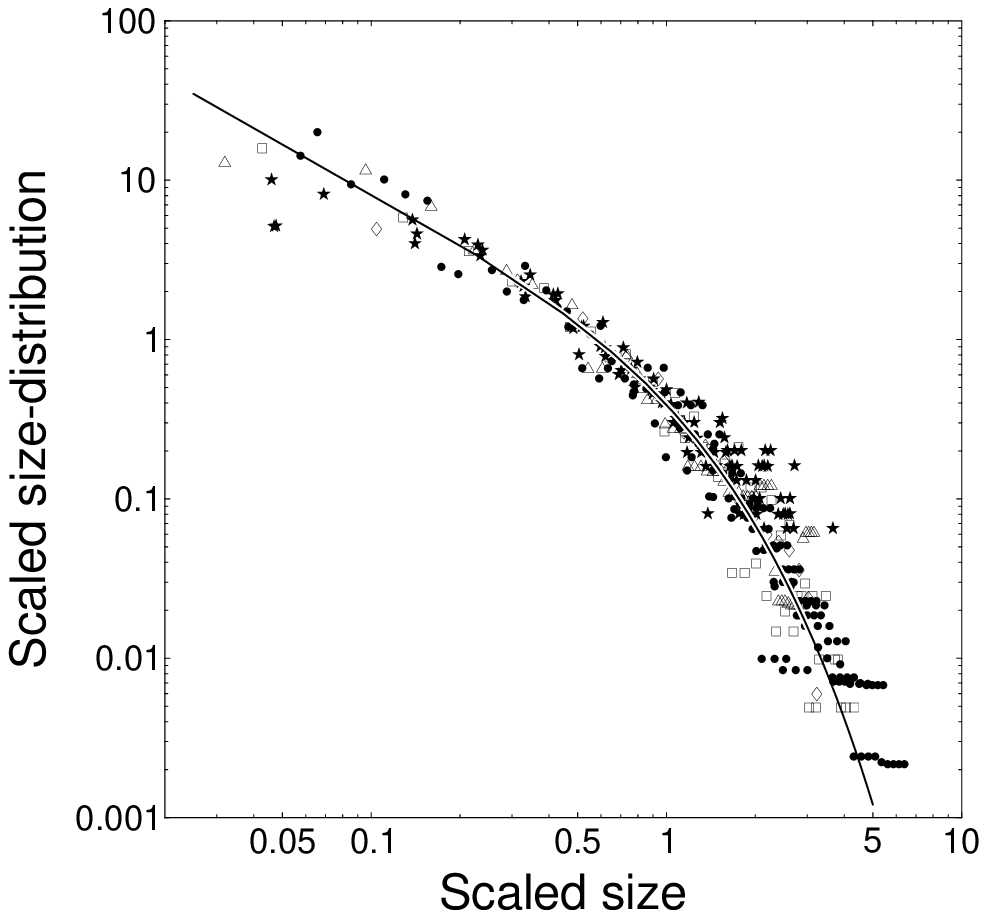}
 \caption{}
 \label{fig:5}
\end{figure}

\begin{figure}[p]
 \centering
 \includegraphics[width=10cm]{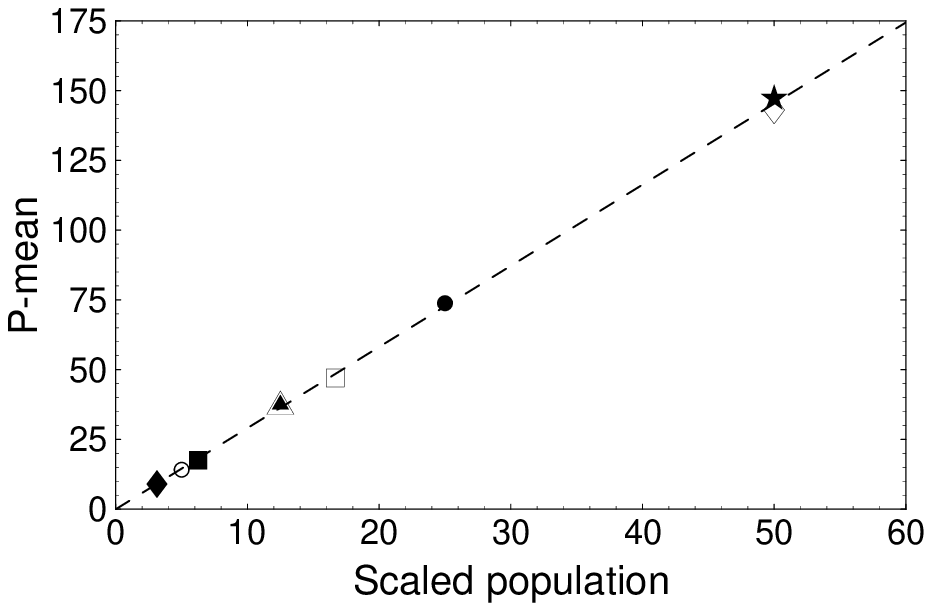}
 \caption{}
 \label{fig:6}
\end{figure}

\begin{figure}[p]
 \centering
 \includegraphics[width=10cm]{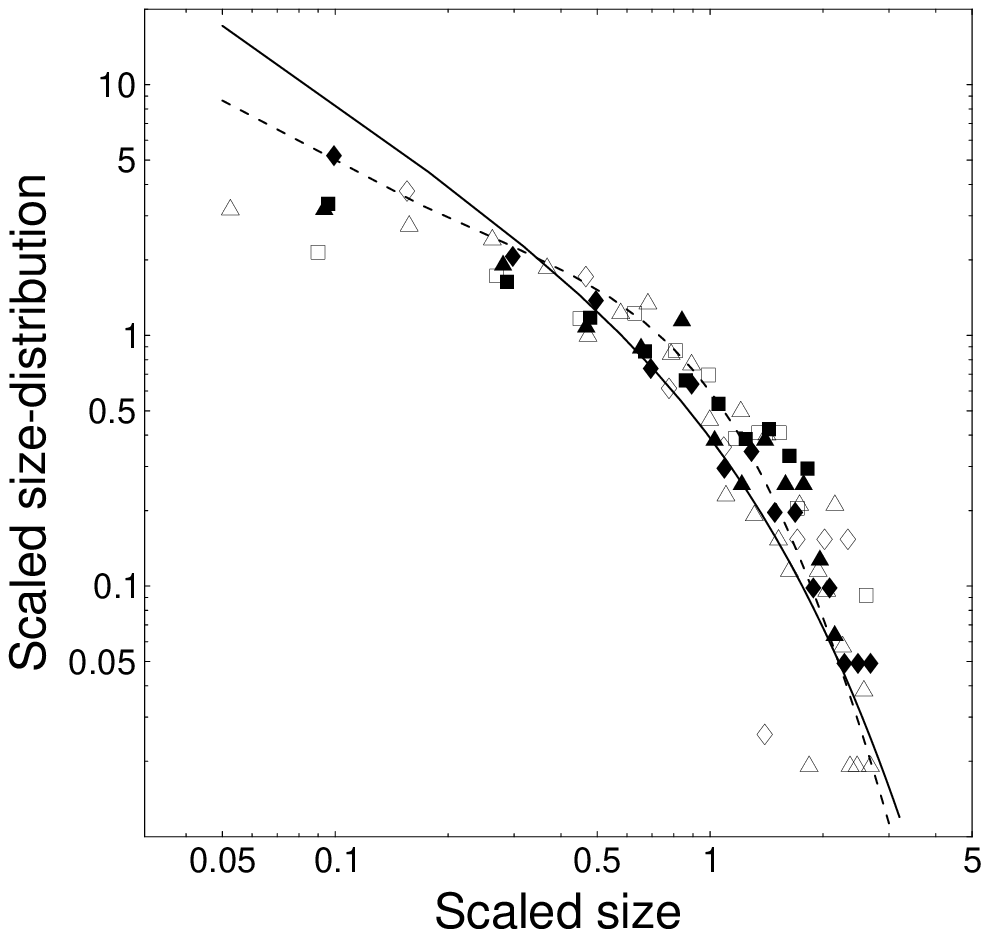}
 \caption{}
 \label{fig:7}
\end{figure}

\begin{figure}[p]
 \centering
 \includegraphics[width=10cm]{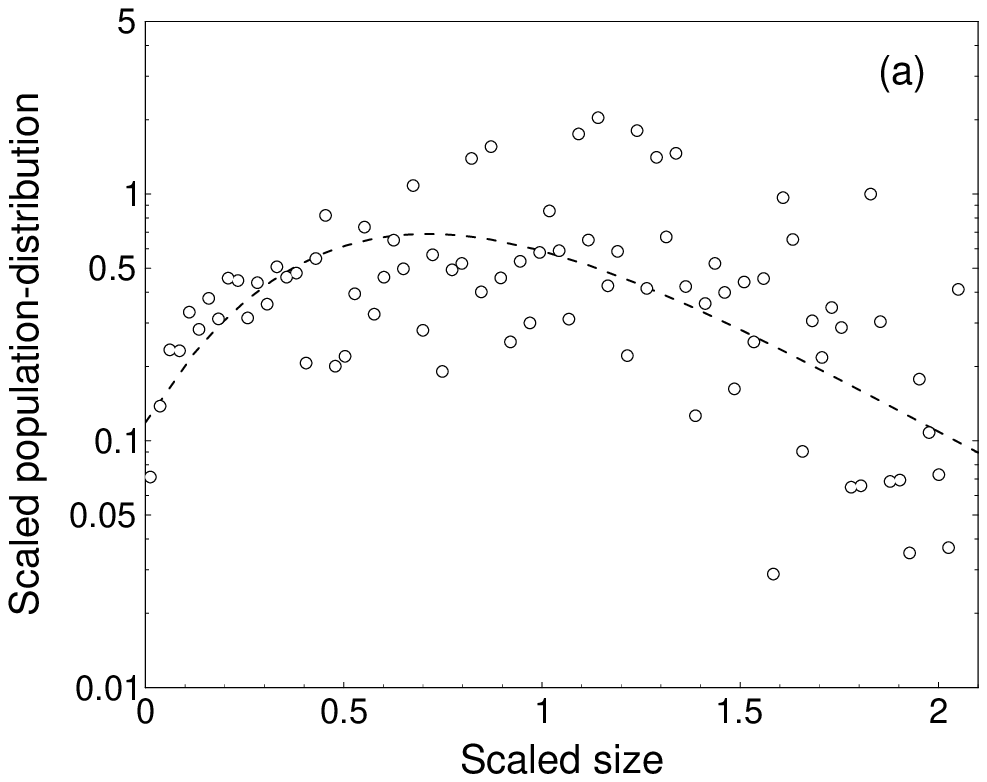}
\\
\vspace*{5mm}
 \includegraphics[width=10cm]{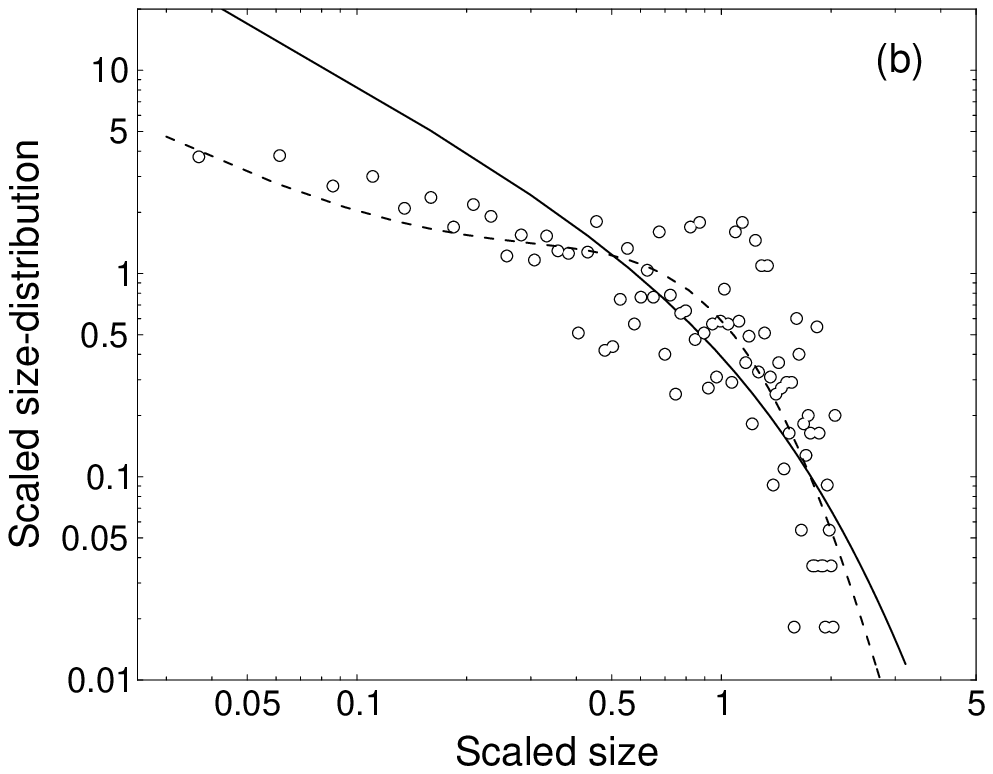}
 \caption{}
 \label{fig:8}
\end{figure}

\begin{itemize}
 \setlength{\itemsep}{10pt}
\item[Fig.1]
Empirical population distribution of pelagic fishes, as shown in Table~\ref{tab1} (except sardinellas).
The scaled distributions
$P_i \langle N \rangle_P$
are plotted against the scaled school sizes
$N_i/\langle N \rangle_P$.
The data are clearly fitted by an exponential decay, suggesting that the distributions are identical,
e.g., $\exp\left({-N/\langle N\rangle_P}\right)$ (broken line).
The proposed model is somewhat more complicated (solid line).
\item[Fig2]
Exponential fluctuations of group-size change (numerical results).
The ordinate is the variance of size change of the group,
$
 \left\langle
 \left(
 \Delta N_{\mbox{\scriptsize A}}
 -\left\langle \Delta N_{\mbox{\scriptsize A}} \right\rangle_N
 \right)^2
 \right\rangle_N
$,
for a finite time interval (single time step of simulation run),
divided by
$\hat{p} \langle N\rangle_P^2$,
where $\left\langle \Delta N_{\mbox{\scriptsize A}} \right\rangle_N$ denotes the expected value of the size change
$\Delta N_{\mbox{\scriptsize A}}$
for a finite time interval, providing that individual ``A'' rides an $N$-sized group.
The abscissa is the group size $N$ divided by $P$-mean.
Simulations show that the group sizes exponentially fluctuate
[solid line represents
 $\sigma^2 (N) \propto \exp\left({N/\langle N\rangle_P}\right)$].
For the simulation parameters, consult Table~\ref{tab2}. 
Note that the variance is scaled based on the fluctuation-dissipation relation
[described later; eqn (\ref{eqn:fluct-dissip})]. 
\item[Fig.3]
Potential induced by the change of variable.
 (a)~Potential $U(x)$ is continuously differentiable (solid line).
 The ordinate is the potential $U(x)$ divided by the breakup rate $p$.
 The abscissa is the reduced size $x$.
 $P$-mean $\langle N\rangle_P$ takes positive values for $x\geq 0$
 and negative values for $x<0$.
 The reduced domain of $x$ is $(-1, 1)$.
 If potential $U(x)$ is continuous
 (broken line showing $U/p$ versus $x/\sqrt{p\langle N\rangle_P^2/D}$),
 the fluctuation-dissipation relation is generalized into eqn (\ref{eqn:g-fluct-dissip}).
 (b)~The group size $N$ as a function of the reduced size $x$.
 The ordinate is the scaled group size
 $N/\langle N\rangle_P$.
\item[Fig.4]
Simulated group-size distribution.
 The scaled distributions
 $W_N \langle N \rangle_P$
 are plotted against the scaled group sizes
 $N/\langle N \rangle_P$.
 The scaled distributions collapse onto a single curve that corresponds to eqn (\ref{w-distr}) with normalization factor eqn (\ref{eqn:normal-factor}).
 The simulation parameters are summarized in Table~\ref{tab2}.
\item[Fig.5]
Empirical school-size distribution of pelagic fishes (the same data sets as Fig.\ref{fig:1}).
 The scaled distributions
 $W_i \langle N \rangle_P$
 are plotted against the scaled school sizes
 $N_i/\langle N \rangle_P$.
 The scaled data collapse onto a single curve that corresponds to eqn (\ref{w-distr}) with normalization factor eqn (\ref{eqn:normal-factor}).
\item[Fig.6]
Plot of $P$-mean against the scaled population size $\Phi/s\hat{p}$, i.e., the ratio of population density to breakup probability (numerical results).
 Parameters are summarized in Table~\ref{tab2}.
 The broken line is a fit with slope $2.907\pm 0.017$.
\item[Fig.7]
Empirical herd-size distribution of mammalian herbivores, as shown in Table~\ref{tab1} (scaled size distributions).
 The solid and broken lines correspond to eqn (\ref{w-distr}) with normalization factor [eqn~(\ref{eqn:normal-factor})] and
 to a least-squares fit of the modified size distribution with
 $x_0^2 = 7.96\pm 1.61$,
 respectively.
\item[Fig.8]
Two kinds of empirical size distributions of sardinellas schools, as shown in Table~\ref{tab1} (scaled size distributions).
 (a)~Population distribution.
 A least-squares fit of the data to eqn (\ref{eqn:g-p-distr}) with
 $x_0^2 = 14.14\pm 1.95$
is shown.
 (b)~School-size distribution.
The solid and broken lines correspond to eqn (\ref{w-distr}) with normalization factor [eqn (\ref{eqn:normal-factor})] and to a fit of the modified school-size distribution, respectively.
\end{itemize}

\newpage
\begin{flushleft}
{\bf Figures}
\end{flushleft}

\begin{itemize}
 \setlength{\itemsep}{5pt}
 \item[Fig.1] Empirical population distribution of pelagic fishes
 \item[Fig.2] Exponential fluctuations of group-size change
 \item[Fig.3] Potential induced by the change of variable
 \item[Fig.4] Simulated group-size distribution
 \item[Fig.5] Empirical school-size distribution of pelagic fishes
 \item[Fig.6] Plot of $P$-mean against the scaled population size
 \item[Fig.7] Empirical herd-size distribution of mammalian herbivores
 \item[Fig.8] Two kinds of empirical size distributions of sardinellas schools
\end{itemize}

\newpage

\newpage

\begin{table}[p]
\caption[database]{
{\bf Species analyzed.}
The ways of estimating school sizes of pelagic fishes were catch per set by a purse seine (in tons) or acoustic surveys (vertical cross-section, vertical thickness or diameter of a school were observed).
Acoustic-survey data are expressed in dimensional size of a school, which can be reduced to the biomass in a school
(Squire, 1978; Anderson, 1981; Misund, 1993; Niwa, 1995; Misund \& Coetzee, 2000):
the school biomass is proportional to the cross-section, the square of thickness or the square of diameter of a school.
The data (distribution density) are given by the set
$\left\{\left.\left(N_i,W_i\right)\right| i=1,2,\ldots,n \right\}$.
$W_i \Delta N$ reads the frequency of group sizes which lie
 within the $i$-th class
$\left[N_i-\Delta N/2, N_i+\Delta N/2\right)$,
where $\Delta N$ denotes the class width
and the $i$-th class mark is given by
$N_i = (i-0.5) \Delta N$.
Data (biomass or reduced number in a group) are normalized as
\(
\sum_{i=1}^n N_i W_i \Delta N = \langle N \rangle_P
\),
where $P$-mean $\langle N \rangle_P$ is defined by eqn (\ref{eqn:pmean}).
The normalization is based on the linear dependence of $P$-mean on the population size~[described later; eqn (\ref{eqn:popsize-Pmean}), Fig.\ref{fig:6}].
}
\vspace*{0.3cm}
\label{tab1}
\begin{center}
\renewcommand{\arraystretch}{1.1}
{
\begin{tabular*}{34.5pc}{llcl}
\noalign{\hrule height0.8pt}
& Species & {$P$-mean$^{\ast}$} & Data sources \\
\noalign{\hrule height0.8pt}
\multicolumn{4}{l}{pelagic fishes}\\
$\square$ & Tropical tuna,$^{\dagger}$ Data from fisheries & 11.69 & \citet{Bonabeau-Dagorn95}\\
$\lozenge$ & Tropical tuna,$^{\dagger}$ Data from fisheries & 4.80 & \citet{Bonabeau-etal99}\\
&\multicolumn{3}{l}{in the vicinity of fish aggregating objects}\\
$\vartriangle$ & {\it Engraulis mordax},$^{\ddagger}$ Acoustic survey &
 15.67 & \citet{Smith70},\\
&&& cited in \citet{Anderson81}\\
$\bullet$ &{\it Sardinops melanosticta}& 6.45--17.36 & Hara (1983, 1985, 1986)\\
&\multicolumn{3}{l}{6 acoustic surveys}\\
$\bigstar$ & {\it Clupea harengus} & 7.24--10.88 & \citet{Reid-etal00}\\
&\multicolumn{3}{l}{4 acoustic surveys}\\
$\circ$ & {\it Sardinellas aurita} and {\it S. maderensis} & 40.73 & \citet{Bonabeau-etal99}\\
&\multicolumn{3}{l}{Catch in the up-welling areas}\\
\hline
\multicolumn{4}{l}{terrestrial herbivores$^{\S}$}\\
$\blacklozenge$ & {\it Gazella thomsoni} & 5.04 & \citet{Wirtz-Lorscher83}\\
$\lozenge$ & {\it Redunca redunca} & 3.22 & \citet{Wirtz-Lorscher83}\\
$\blacksquare$ & {\it Kobus ellipsiprymnus} & 5.22 & \citet{Wirtz-Lorscher83}\\
$\square$ & {\it Bison bison} & 5.56 & \citet{Lott-Minta83}\\
$\blacktriangle$ & {\it Syncerus caffer} & 5.35 & \citet{Sinclair77}\\
$\vartriangle$ & {\it Ovis canadensis} & 9.53 & \citet{Hansen80}\\
\noalign{\hrule height0.8pt}
\multicolumn{4}{l}{$^{\ast}$A unit size is $(N_1+N_2)/2$, which may contain a certain number of individuals.}\\
\multicolumn{4}{l}{$^{\dagger}$Three species ({\it Thunnus
 albacares}, {\it Katsuwonus pelamis}, and {\it Thunnus obesus})}\\
\multicolumn{4}{l}{are mixed.}\\
\multicolumn{4}{l}{$^{\ddagger}$possibly including {\it
 Trachurus symmetricus}, {\it Sarda chiliensis}, {\it Scomber
 japonicus},}\\
\multicolumn{4}{l}{and {\it Sardinops sagax}.}\\
\multicolumn{4}{l}{$^{\S}$Data are cited in \citet{Okubo86}.}\\
\end{tabular*}
}
\end{center}
\end{table}

\newpage

\begin{table}[p]
\caption[simulation]{
 {\bf Parameters used in simulations.}
Simulations have been performed with the coarse-grained zones of $s=2^{18}$ sites, simulation run $=2^{21}$ time steps, and parameters summarized in the table.
The splitting probability at each time step of simulation run is defined by eqn (\ref{eqn:split-prob}).
The frequency of the amount of $N$-sized groups ($N=1,2,\ldots,\Phi$) at
 the last of simulation run is represented by $W_N$ in Fig.\ref{fig:4},
where the frequency distribution is normalized as
$\sum_{N=1}^\Phi N W_N = \langle N\rangle_P$.
$P$-mean sizes were computed from simulation results.
}
\vspace*{0.3cm}
\label{tab2}
\begin{center}
\renewcommand{\arraystretch}{1.1}
{
\begin{tabular*}{21.8pc}{lccc}
\noalign{\hrule height0.8pt}
& breakup probability $\hat{p}$ & population $\Phi$ & $P$-mean\\
\noalign{\hrule height0.8pt}
$\blacklozenge$ & 0.02 & $2^{14}$ & 9.00\\
$\blacksquare$ & 0.02 & $2^{15}$ & 17.54\\
$\blacktriangle$ &0.02 & $2^{16}$ & 37.52\\
$\bullet$ & 0.02 & $2^{17}$ & 73.80\\
$\bigstar$ & 0.02 & $2^{18}$ & 147.27\\
\hline
$\lozenge$ & 0.01 & $2^{17}$ & 143.03\\
$\square$ & 0.03 & $2^{17}$ & 47.01\\
$\vartriangle$ & 0.04 & $2^{17}$ & 36.77\\
$\circ$ & 0.1 & $2^{17}$ & 14.14\\
\noalign{\hrule height0.8pt}
\end{tabular*}
}
\end{center}
\end{table}

\end{document}